\documentclass[12pt]{article}
\usepackage[T1]{fontenc}
\usepackage[francais,english]{babel}
\usepackage[dvips]{graphicx}
\usepackage{amsfonts}
\usepackage{latexsym}
\usepackage{mathptmx}
\usepackage{amsmath}
\usepackage{lscape}
\usepackage{longtable}
\usepackage{lscape}
\usepackage{rotating}
\usepackage{longtable}
\usepackage{setspace}
\usepackage{anysize}

\marginsize{.9in}{.9in}{1.5in}{1.5in}

\def\picture #1 by #2 (#3){\dimen0 = \hsize
\advance\dimen0 by -#1 \divide\dimen0 by 2 \hskip
\dimen0 \vbox to #2{\hrule width #1 height 0pt depth 0pt
\vfill \special{picture #3}}}

\def \R{{\mathbb R}}

\font\ninerm=cmr9
\long\outer\def\abstract#1{\bigskip\vbox{\noindent\ninerm
\baselineskip=10pt#1}\nobreak\bigskip}
\catcode`\â= \active \def â{\^a}
\catcode`\ê= \active \def ê{\^e}
\catcode`\ô= \active \def ô{\^o}
\catcode`\û= \active \def û{\^u}
\catcode`\î= \active \def î{{\^\i}}
\catcode`\é= \active \def é{\'e}
\catcode`\è= \active \def è{\`e}
\catcode`\à= \active \def à{\`a}
\catcode`\ù= \active \def ù{\`u}
\catcode`\ç= \active \def ç{\c {c}}
\catcode`\ï= \active \def ï{{\"\i}}

\newcount\numero
\def\exo#1{\advance\numero by 1\bigskip
{\noindent\tenbf #1\the\numero. }}
\def\frac#1#2{{#1\over #2}}
\numberwithin{equation}{section}

\begin{document}

\title{On the Necessity of Five Risk Measures}

\author{Dominique Guégan\footnote{Paris School of Economics, MSE - CES, Université Paris1 Panthéon-Sorbonne, 106 boulevard de l'hopital, 75013 Paris, France, e-mail: dominique.guegan@univ-paris1.fr}
, Wayne Tarrant\footnote{Department of Mathematics, Wingate University, Wingate, NC 28174, USA, e-mail: w.tarrant@wingate.edu}}

\maketitle

\begin{abstract}

\noindent \textit{Abstract}:  The banking systems that deal with risk management depend on underlying risk measures. Following the Basel II accord, there are two separate methods by which banks may determine their capital requirement. The Value at Risk measure plays an important role in computing the capital for both approaches. In this paper we analyze the errors produced by using this measure. We discuss other measures, demonstrating their strengths and shortcomings. We give examples, showing the need for the information from multiple risk measures in order to determine a bank's loss distribution. We conclude by suggesting a regulatory requirement of multiple risk measures being reported by banks, giving specific recommendations.\\



\end{abstract}

\vfil \eject

\noindent In the middle of 2008, the worldwide banking system seemed to start unraveling. Some consumers failed to pay the payments on their mortgages or other loans. This led to default on several collateralized debt obligations (CDOs), triggering the need for payment of the default leg of many credit default swaps (CDSs). A fear of counterparty risk seized the banking industry, not knowing if the parties who had sold the insurance against the default on the CDOs would be able to make the contracted payments. This led to a ``credit seizure'' whereby banks stopped lending money as freely as they once had, both to consumers and to other banking entities.\\

\noindent Many people have asked the question of how such a series of events took place. Although the reasons are complicated and a political hot button topic, some citizens have come to learn more about worldwide banking systems and their regulators. Many are calling for renewed regulation of banks and similar enterprises. But in order to craft meaningful banking laws, we must first understand what laws have been in place and why we have had problems such as those described above in spite of the laws that were in effect.\\

\noindent The banking laws that deal with risk management are based upon underlying risk measures that are used in order to determine the capital requirement for banks.  In the following, we consider the strengths and weaknesses of the most used risk measure and then consider other risk measures. We will discuss the limitations of each of these individual risk measures and then show that at least five risk measures are necessary for distinguishing among different types of risk profiles when determining capital requirements for banks.\\

\noindent We note that this approach is a new one because, in general, researchers and practitioners limit themselves to the study or use of a single measure or sometimes two measures. (See, for example, Artzner et al. (1997), Yamai and Yoshiba (2002), and Caillault and Gu\'{e}gan (2009), among others.) Here we insist upon the use of a larger number of measures of risk, which are easy enough to compute in practice. We show that this information gives a better understanding of the risk incurred and more insight into deciding the best risk management strategy for banks. This approach is also novel because many researchers view risk measures solely as a way to compute bank capital reserve requirements. We suggest that each piece of information that a bank releases through its reporting of a risk measure may be used to give a fuller picture of a bank's loss distribution. This should be valuable to investors and of utmost importance to regulatory bodies. \\

\noindent The paper is organized as follows. In Section one we have recalled the notion of risk measures. Section two is devoted to a discussion of regulation in the banking industry. Section three focuses on the problems related to the use of Value at Risk and discusses the notion of coherent risk measures. Section four presents different scenarios using more than one risk measure, finally proposing at least five measures. The final section summarizes our recommendations and speaks of future directions.

\section*{I. Some definitions}

\noindent To understand international capital requirement regulations without interrupting the flow, it is necessary to specify the notion of risks with which we work or that we wish to measure. There is little wonder that people are confused by what the term ``risk'' ought to mean. For instance, the website businessdictionary.com defines risk in seventeen different ``general categories''. For an illustration of the complexity of the problem, we now recall this definition: \\

\noindent ``Risk is the probability that an actual return on an investment will be lower than the expected return. Financial risk is divided into the following general categories: (1) Basis risk: Changes in interest rates will cause interest-bearing liabilities (deposits) to reprice at a rate higher than that of the interest-bearing assets (loans). (2) Capital risk: Losses from un-recovered loans will affect the financial institution's capital base and may necessitate floating of a new stock (share) issue. (3) Country risk: Economic and political changes in a foreign country will affect loan-repayments from debtors. (4) Default risk: Borrowers will not be able to repay principal and interest as arranged (also called credit risk). (5) Delivery risk: Buyer or seller of a financial instrument or foreign currency will not be able to meet associated delivery obligations on their maturity. (6) Economic risk: Changes in the state of economy will impair the debtors' ability to pay or the potential borrower's ability to borrow. (7) Exchange rate risk: Appreciation or depreciation of a currency will result in a loss or a naked-position. (8) Interest rate risk: Decline in net interest income will result from changes in relationship between interest income and interest expense. (9) Liquidity risk: There will not be enough cash and/or cash-equivalents to meet the needs of depositors and borrowers. (10) Operations risk: Failure of data processing equipment will prevent the bank from maintaining its critical operations to the customers' satisfaction. (11) Payment system risk: Payment system of a major bank will malfunction and will hinder its payments. (12) Political risk: Political changes in a debtor's country will jeopardize debt-service payments. (13) Refinancing risk: It will not be possible to refinance maturing liabilities (deposits) when they fall due, at economic cost and terms. (14) Reinvestment risk: It will not be possible to reinvest interest-earning assets (loans) at current market rates. (15) Settlement risk: Failure of a major bank will result in a chain-reaction reducing other banks' ability to honor payment commitments. (16) Sovereign risk: Local or foreign debtor-government will refuse to honor its debt obligations on their due date. (17) Underwriting risk: New issue of securities underwritten by the institution will not be sold or its market price will drop.''\\

\noindent Each of these seventeen definitions are relevant to the situations of banks and securities firms. The unifying theme for each of the above definitions is that the situation of risk requires both uncertainty and exposure. If a company already knows that a loan will default, there is no uncertainty and thus no risk. And if the bank decides not to loan to a business that is considered likely to default, there is also no risk for that bank as the bank has no exposure to the possibility of loss. Thus, Merriam-Webster (2003) defines risk as the possibility of loss or injury. This is certainly consistent with the concepts we have set forth.\\

\noindent So now one goal could be to have a single risk measure that will account for all the risk that a bank or securities firm might encounter. Some have objected to the risk measure being a single number, but there is some support for this idea. Investing is always a binary decision- either one invests or one chooses not to invest. Thus the argument is that, given a single number, one should have enough information to decide whether to invest or not. There have been some general agreements about the kinds of properties that such a risk measure ought to possess. These agreements must be acknowledged as being assumptions, but they have provided the definition that is now universal.\\

\noindent We recall the universally accepted definition of a risk measure (Artzner et al., 1997). Let $X$ be a random variable. Then $\rho$ is a risk measure if it satisfies the following properties:
\begin{itemize}
\item (Monotonicity) if $X \ge 0$, then $\rho(X) \le 0$
\item (Positive Homogeneity) $\rho(\alpha X) = \alpha \rho(X) \ \forall \ \alpha \ge 0$
\item (Translation Invariance) $\rho(X + a) = \rho(X) - a \ \forall \ a \in \R$.
\end{itemize}

\noindent The idea behind this definition is that a positive number implies that one is at risk for losing capital and should have that positive number of a cash balance on hand to offset this potential loss. A negative number would say that the company has enough capital to take on more risk or to return some of its cash to other operations or to its shareholders. The monotonicity property states that an investment that always has positive payoff gives the company the ability to take on more risk. Positive homogeneity implies that multiplying your investment by $k$ times gives you a risk of a loss that is $k$ times larger. Translation invariance implies that a company holding $a$ in cash lowers its measure of risk by $a$.\\

\noindent The most well-known measure seems to be the Value at Risk (VaR). In order to define VaR, we must first recall some basic concepts.
Let $X$ be a random variable and $\alpha \in \lbrack 0,1 \rbrack $.
\begin{itemize}
\item  $q$ is called an \bf{$\alpha$-quantile} \rm if $Pr \lbrack X < q \rbrack \le 1 - \alpha \le Pr \lbrack X \le q \rbrack$,
\item the largest \bf{$\alpha$-quantile} \rm is $q_{\alpha}(X) = inf \lbrace x | Pr \lbrack X \le x \rbrack > 1 - \alpha \rbrace$, and
\item the smallest \bf{$\alpha$-quantile} \rm is $q_{\alpha}^- = inf \lbrace x | Pr \lbrack X \le x \rbrack \ge 1 - \alpha \rbrace$.
\end{itemize}

\noindent It is easy to show that $q_{\alpha} \ge q_{\alpha}^-$ and that $q$ is an $\alpha$-quantile if and only if $q_{\alpha}^- \le q \le q_{\alpha}$.\\

\noindent Given a position $X$ and a number $\alpha \in \lbrack 0,1 \rbrack$, we define the $\alpha$-Value at Risk,  $VaR_{\alpha}(X)$, by $VaR_{\alpha}(X) = -q_{\alpha}(X)$. We call $X$ $\alpha$-VaR acceptable  if $VaR_{\alpha}(X) \le 0$ or, equivalently, if $q_{\alpha}(X) \ge 0$. The $\alpha$-VaR can be seen as the amount of cash that a firm needs in order to make the probability of that firm going bankrupt to be equal to $\alpha$. This leads to the following property of VaR: $VaR_{\alpha}(X + VaR_{\alpha}(X)) = 0$. This states that one may offset the risk of an investment by having an amount of cash on hand equal to the Value at Risk inherent in holding the asset.\\

\noindent The Value at Risk has historically been the most important of the risk measures, as it is the one used by the Basel Committee on Banking Supervision in order to determine capital requirements for banks. Now, we briefly discuss the history of the Basel Committee, and then we will delineate issues with the Value at Risk.

\section*{II. The Basel Committee on Banking Supervision}

\noindent The Basel Committee has its origins in the Bank Herstatt issue of 1974. On 26 June 1974 German regulators closed Bank Herstatt and forced its liquidation. That morning several banks had made payment to Bank Herstatt in Deutsche Marks in Frankfurt in exchange for United States Dollars that were to be delivered in New York later in the day and not simultaneously only due to time zone differences. Bank Herstatt was closed between the times of the two payments, meaning that the counterparty banks did not receive their US dollars in New York.\\

\noindent Because of this debacle, and due to the issue of jurisdictions for such a situation, the G-10 countries formed a committee under the Bank of International Settlements,  (BCBS, 1988). This Basel Committee on Banking Supervision consists of representatives from central banks and from regulatory agencies. The focus of this committee continues to grow and change, but it has historically tried to accomplish the following goals: to define roles of regulators in cross-jurisdictional situations; to ensure that international banks or bank holding companies do not escape comprehensive supervision by a native regulatory authority; to promote uniform capital requirements so that banks from different countries may compete with one another on a ``level playing field''.\\

\noindent It is the third goal that we will address at length. It is important to note, though, that the Basel Committee has no legislative authority, but the participating countries are honor bound to use its recommendations in their own regulations. However, the committee seems to understand and tolerate some difference in how each country implements recommendations due to historical or cultural differences in each country's banking community.\\

\noindent In trying to address the goal of a ``level playing field,'' the Basel Committee proposed minimal capital requirements for banks in 1988. In 1992 these requirements became law in the G-10 countries, with the notable exception of Japan, which took a longer period of transition to these requirements due to the nature of the banking culture in that country. These regulations have come to be called the 1988 Basel Accord or Basel I (BCBS 1988).\\

\noindent This 1988 Basel Accord involved banking in the sense of taking deposits and lending (which is known as commercial banking under US law), and the 1988 Basel Accord's thrust was at credit risk. Under these requirements, a bank would calculate its capital and its credit risk and then have to meet the stipulation that
$$\frac{\texttt{capital}}{\texttt{credit risk}} \ge 8\%$$

\noindent This is often called an 8\% capital requirement. In addition, the 1988 Basel Accord distinguished two different types, or tiers, of capital. Tier 1 capital is comprised of the book value of common stock, non-cumulative perpetual preferred stock, and published reserves from post-tax retained earnings. Tier 2 capital is assumed to be of lower quality and includes general loan loss reserves, long-term subordinated debt, and cumulative and/or redeemable preferred stock. Under the 1988 Basel Accord, up to 50\% of a bank's stated capital (for the 8\% capital requirement) could be tier 2 capital.\\

\noindent The credit risk was calculated in a somewhat more complicated manner as a sum of risk-weighted asset values. Although the Basel Committee wanted to make a simple system and only used five different weights, there are still many asset classes to delineate. The Committee also left some of the supervision up to local authorities in that the class of claims similar to municipal bonds was given one of four weights at the discretion of the country's regulatory agency. The specifications on risk weights are as follows: \\

\begin{itemize}
\item 0\%:
\begin{enumerate}
\item Cash (includes, at national discretion, gold bullion held in own vaults or on an allocated basis to the extent backed by bullion liabilities)
\item Claims on central governments and central banks denominated in national currency and funded in that currency
\item Other claims on Organization of Economic Cooperation and Development (OECD) central governments and central banks
\item  Claims collateralized by cash of OECD central-government
securities or guaranteed by OECD central governments
\end{enumerate}
\item 0, 10, 20 or 50\% (at national discretion):
\begin{enumerate}
\item Claims on domestic public-sector entities, excluding central
government, and loans guaranteed by such entities
\end{enumerate}
\item 20\%
\begin{enumerate}
\item Claims on multilateral development banks (International Bank for Reconstruction and Development, Inter-American Development Bank, Asian Development Bank, African Development Bank, European Investment Bank) and claims guaranteed by, or collateralized by securities issued by such banks
\item  Claims on banks incorporated in the OECD and loans guaranteed
by OECD incorporated banks
\item  Claims on banks incorporated in countries outside the OECD with a
residual maturity of up to one year and loans with a residual
maturity of up to one year guaranteed by banks incorporated in
countries outside the OECD
\item Claims on non-domestic OECD public-sector entities, excluding
central government, and loans guaranteed by such entities
\item Cash items in process of collection
\end{enumerate}
\item 50\%
\begin{enumerate}
\item Loans fully secured by mortgage on residential property that is or
will be occupied by the borrower or that is rented
\end{enumerate}
\item 100\%
\begin{enumerate}
\item Claims on the private sector
\item Claims on banks incorporated outside the OECD with a residual
maturity of over one year
\item  Claims on central governments outside the OECD (unless
denominated in national currency - and funded in that currency -
see above)
\item Claims on commercial companies owned by the public sector
\item  Premises, plant and equipment and other fixed assets
\item  Real estate and other investments (including non-consolidated
investment participation in other companies)
\item  Capital instruments issued by other banks (unless deducted from
capital)
\item  All other assets
\end{enumerate}
\end{itemize}

\noindent In the early 1990s, the Basel Committee decided that it needed to account for market risk. In April of 1993 the Basel Committee put forth a series of amendments to the 1988 Accord. These amendments were basically intended to give capital requirements for banks' market risk. Banks would be required to identify their trading book and hold capital to offset the risk associated with this trading book. Capital charges for offsetting the market risk of the trading book would be based upon a 10-day 95\% VaR measure. The Committee also added a category of Tier 3 capital, which was short-term subordinated debt. However, this Tier 3 capital could only be used to cover market risk. Now, the inequality that banks had to satisfy was:
$$\frac{\texttt{capital}}{\texttt{credit risk} + \texttt{market risk}} \ge 8\%$$

\noindent After a number of comments on these revisions, the Basel Committee released a revised proposal in April 1995. There were a number of changes to the 1993 amendments, most notably that the Basel Committee would now allow banks to use either the previously announced VaR-based measures or individual proprietary VaR measures (the Internal Ratings Based or IRB Approach) for computing capital requirements. The proposals were adopted in 1996 and quickly became known as the 1996 amendment, BCBS (1996). This went into effect in 1998.\\

\noindent Later, it was recognized that the Accord's treatment of credit risk had problems. The weights provided an incentive for banks to hold the debt of the G-10 governments, which paid very small dividends to the banks. This fact posed a glaring conflict of interest because the G-10 governments were the ones with regulatory oversight of the banks. Corporate debt was much more profitable, but the banks were charged at the 100\% rate for holding this debt. Banks quickly realized that all corporate debt was weighted equally, while some corporate debt was much more risky and thus supplied a larger reward. Banks came to hold a lot of risky corporate debt and less high quality corporate debt. This practice became known as regulatory arbitrage. \\

\noindent However, the 1988 Accord's system of risk weights also failed to address a burgeoning market. Banks used both credit derivatives and securitizations to take advantage of the Accord. In addition there was outright fraud being perpetrated in banks, either internally or by borrowers from the banks. The original Basel Accord and its amendments failed to speak to these issues of risk. A new accord would have to address such operational risk.\\

\noindent In 1999 the Basel Committee proposed a new Accord, now known as Basel II. Basel II professes to be based on three pillars: minimum capital requirements, supervisory review, and market discipline. Basel II (BCBS, 2005) replaces the original treatment of credit risk and requires capital to offset operational risk, thus making the new inequality:
$$\frac{\texttt{capital}}{\texttt{credit risk} + \texttt{market risk} + \texttt{operational risk}} \ge 8\%$$

\noindent However, there is no change to the computation of bank capital (the three tiered system) or to defining market risk. The banks are still allowed to use proprietary VaR-based measures by choosing the IRB, or they can opt to use the standardized measure. They also now have three choices for computing credit risk and three more choices for computing operational risk. Basel II went into effect in December 2006, but it has not been followed as widely as the previous accords. For instance only ten of the largest banks in the US are subject to Basel II (as the US believes its own Securities and Exchange Commission has more stringent rules that Basel II), while another ten have the option to opt-in to such regulation.\\

\noindent The Basel Committee did take some steps to address the risk weights on sovereign debts. It was recognized that the differing risk weights on the debts of OECD and non-OECD countries was arbitrary at best. Nations outside of the OECD like South Africa and Brazil would often have lower probabilities of default than countries like Greece or Spain, both of which belong to the OECD. So in the Basel II accords, countries' debts are given risk weights based upon the ratings of credit agencies. A rating from AAA to AA- gets a risk weight of 0\%, from A+ to A- earns a 20\% risk weight, from BBB+ to BBB- garners a 50\% risk weight, from BB+ to B- (or unrated) achieves a country a risk weight of 100\%, and anything in the junk category (i.e. below B-) implies a risk weight of 150\%. \\

\noindent What is most notable about all the changes is that under the standardized approach the VaR-based measures and the discrete risk weights remain. Central to understanding why banks have struggled so much lately is understanding their risk management, and in order to understand risk management, we must understand Value at Risk and investigate the reliability of the risk weights.

\section*{III. Value at Risk and coherent measures}
\subsection*{A. Problems with Value at Risk}

\noindent There are two major problems cited with Value at Risk as a risk measure.\\

\noindent  The issue that is easier to describe is the fact that Value at Risk does not address how large of a bankruptcy a firm will experience. This risk measure only tells that a business will be bankrupt with a 5\% chance if it doesn't have on account a sum of money equal to the 95\%-VaR.
As an example, say that a firm holds an asset that initially had no cost. That asset has a 95\% chance of paying the company US\$2 million and a 5\% chance of giving the company a loss of US\$1 million. By the above definition, we find that the 95\%-VaR is US\$1 million.
Now assume that the firm holds another asset, again without initial cost. This asset again has a 95\% chance of giving the company a profit of US\$2 million, with a 1\% chance of a loss of US\$1million and a 4\% chance of a loss of US\$1 billion. This asset also has a 95\%-VaR of US\$1 million, but it is very clear that the second asset is much more risky for the firm to hold.\\

\noindent One can claim that such an example is contrived or that a different value of $\alpha$ might eradicate this problem. First, this issue has already appeared in the problems with paying off credit default swaps. Second, the mathematics can be easily shifted to account for a 99\%-VaR or a 99.9\%-VaR or whatever level one might choose. Considering that returns on stocks have been shown to have the fatter tailed Pareto distributions (Mandelbrot (1963), Fama (1965)), very large losses are possible with small probability. This example of Pareto distributions could be used to address both the objections of contrivance and of scaling of $\alpha$ with the same class of examples.\\

\noindent The second problem critics cite is that VaR does not account for diversification effects. It is widely held that owning different types of investment vehicles will help to alleviate some of the risk that holding only one investment would produce. Markowitz (1952) discusses diversifying away specific risk, that risk that is unique to a given holding. He acknowledges that one cannot alleviate the market risk that is common to all holdings. Markowitz's calculations show that holding a sufficiently large number of assets, chosen with appropriate covariance, diversifies away specific risk.\\

\noindent As an example of this problem, consider the case where a bank has made two \$1 million loans and one \$2 million loan, each with a 0.04 probability of default and all pairwise independent. Then the 95\%-VaR for each loan is \$0. Thus, if we construct a portfolio consisting solely of the \$2 million loan, we must have a 95\%-VaR of \$0. If we instead choose diversification and make our portfolio out of the two independent \$1 million loans, something paradoxical happens. The probability of both loans defaulting is 0.0016, but the probability of exactly one loan defaulting is 0.0768. This implies that the 95\%-VaR of our diversified portfolio is \$1 million. Thus, VaR does not favor diversification in some instances.\\

\noindent This should lead us to question whether we ought to favor diversification or whether diversification is a property that is not meaningful to us. For instance, at www.glyn\-hylton.com, the following scenario is proposed:
Imagine you are stranded on a desert island. For fresh water, there are three natural springs, but it is possible one or more have been poisoned. To minimize your risk, what is your optimal strategy for drinking from the springs? You might: select one of the three springs at random and drink exclusively from it or select two of the springs at random and drink exclusively from them, or drink from all three springs.
Diversification would say that we would reduce our risk by drinking from each of the three streams. Yet common sense says that if one spring is poisoned, we should not take a single drink from it for fear of being killed by ingesting even a small amount of the poison. So, if we determine from the first drink from one spring that it is not poisoned, we have no incentive to diversify away our risk by drinking from the others. Instead, this would only add to our risk of being poisoned. One can easily see this analogy applied to the area of risky assets.\\

\noindent  So, is diversification always the best policy? If each of our assets has some probability of default, is it better to spread the risk of loss over several loans, choosing a larger probability of taking a small loss? Or does it make more sense to hold only one asset, taking a small probability of losing everything? These are questions of risk aversion that each bank and each individual investor will need to answer depending on his or her or its own risk preferences. Investment professionals have decided that the former is preferable to the potential of losing their entire investment. And some regulating agencies have agreed with this judgment, though the Basel Committee has implicitly disagreed, whether intentionally or not. Thus, we will pursue ways in which we might reconcile these differences and determine if there is some risk measure that might respect the desirability of diversification.

\subsection*{B. Coherent Risk Measures}

In 1997 Artzner, Delbaen, Eber, and Heath (1997) (henceforth ADEH) proposed four desirable properties that measures of risk should have. In addition to the properties of translation invariance, positive homogeneity, and monotonicity, these authors added the property of subadditivity, i.e., using the notation from before,
$$
{\rm if} \; \; X_1, X_2\;  {\rm are}\;  {\rm random}\; {\rm variables,}\; {\rm then} \; \;  \rho(X_1 + X_2) \le \rho(X_1) + \rho(X_2).
$$
\noindent This is an attempt to respect the perceived value of portfolio diversification. They chose to call any measure that satisfied all four of the properties by the name ``coherent risk measure''. Of course this was an intelligent choice of names for their class of measures, as nobody would want to be accused of using a measure that is incoherent.\\

\noindent ADEH (1997) start by defining a set of acceptable positions. These are positions that some sort of supervisor, such as a regulator or an exchange's clearing firm, decides have an appropriate level of risk for the firm. They then propose that a firm must make choices when faced with an unacceptable position. Either the firm may alter the position or decide to offset the unacceptable position by having cash on hand to account for the possibility of a loss. Should the firm decide to hold the unacceptable position, a risk measure will be the ``distance'' away from acceptable that their position measures.\\

\noindent ADEH (1997) then go about showing links between acceptance sets and measures of risk. Their acceptance sets are those for which the risk measure $\rho$ takes a negative value on the set. Thus, acceptance sets are those for which no additional capital is needed to offset the risk that is being taken by holding the given position.\\

\noindent They next demonstrate that VaR is incoherent and also that VaR fails to recognize when there is a concentration of risks. This is just as we have shown above, although their examples are slightly more complicated and less transparent. The main conclusions of these authors  to reject Value at Risk as a good risk measure  lies on the following two reasons:
\begin{itemize}
\item  Value at Risk does not behave nicely with respect to addition of risks, even independent ones, creating severe aggregation problems.
\item the use of Value at Risk does not encourage and, indeed, sometimes prohibits diversification, because Value at Risk does not take into account the economic consequences of the events the probabilities of which it controls.
\end{itemize}

\noindent The authors (ADEH, 1997) then give the above definition of a coherent risk measure. They prove that any coherent risk measure arises as the supremum of the expected negative of final net worth for some collection of ``generalized scenarios'' or probability measures on states of the world. They point out that ``Casualty actuaries have been working long at computing pure premium for policies with deductible, using the conditional average of claim size, given that the claim exceeds the deductible, (Hogg and Klugman, 1984). In the same manner, reinsurance treaties
have involved the conditional distribution of a claim for a policy (or of the total
claim for a portfolio of policies), given that it is above the ceding insurer's retention
level. In order to tackle the question of \it{how} \rm bad is bad, which is not addressed by
the Value at Risk measurement, some actuaries (Bassi, Embrechts and Kafetzaki, 1996) have first identified
the deductible (or retention level) with the quantile used in the field of financial
risk measurement.'' \\

\noindent We now introduce the notion of tail conditional expectation.  Given a base probability
measure P on a probability space $\Omega$ and a level $\alpha$, the tail conditional expectation is the measure of risk defined by
\begin{equation} \label{tce}
TCE_{\alpha}(X) = -E_P [X | X \le - VaR_{\alpha}(X)]
\end{equation}

\noindent This concept is variously called the expected shortfall, expected tail loss, conditional value at risk, loss given default, mean excess loss or mean shortfall by several authors in the literature, Rockafellar  and Uryasev, (2002).
(ADEH) are able to show that coherent risk measures in general (and so their tail conditional expectation in particular) correspond to convex closed sets which include the positive orthant and are in fact homogeneous cones. They give several classifications, showing that coherent risk measures are in correspondence with the measures that arise from looking at the ``generalized'' scenarios that banks may face in order to construct and maintain acceptable positions. In mathematical terms, these are measures on probability spaces. \\

\noindent Finally, we can remark that for each risk $X$ one has the equality
$$VaR_{\alpha}(X) = inf \lbrace \rho(X) | \rho \; \; {\rm coherent}, \; \; \rho \ge VaR_{\alpha}(X) \rbrace.$$
Thus, knowing that
 more restrictive measures are available to them, the question is why regulators would use the Value at Risk, noting that no known organized exchanges use VaR as the basis of risk measurement for margin requirements. ADEH (1997) immediately answer their own question, with a quote from Stulz (1996), ``Regulators like Value at Risk because they can regulate it.'' \\

 \noindent We will look at the problems that will arise from such an attitude.

\section*{IV. How many measures do we need for a robust risk management strategy?}

We note that in the entirety of section four we consider a single timeframe for all the risk measures under consideration. That is to say that all calculations are assumed to be over the identical historical timeframe. For example we might consider a 100-day 99\% VaR and a 100-day 99\% TCE, but in this section we will not consider a 100-day 99\% VaR and a 50-day 99\% TCE.

\subsection*{A. Insufficiency of a single measure}

\noindent Here, we discuss the weakness of any single measure of risk for risk management strategy. \\

\noindent We start by noting that each of the graphs that are shown in this and the following sections represent the final 5\% of the loss probability density function, i.e. if $f(x)$ is the pdf of a loss distribution $X$ that we are considering, then we must have that $\int_c^d f(x) dx = 0.05$. In this case $\lbrack c,d \rbrack$ is the region of the graph that we have shown. We point out this region in Figure 1. In the following figures, we will only see the region $\lbrack c,d \rbrack$. There is an understood assumption that $f(x) = 0$ for all $x > d$. Under this notation, the 95\%-VaR($X$) = $c$ and the 95\%-TCE($X$) = $\frac{1}{.05} \int_c^d x f(x) dx $. We note that the probability density function of the loss is just the negative of the probability density function for the returns.\\

\begin{figure}[h!]
\centering%
\includegraphics{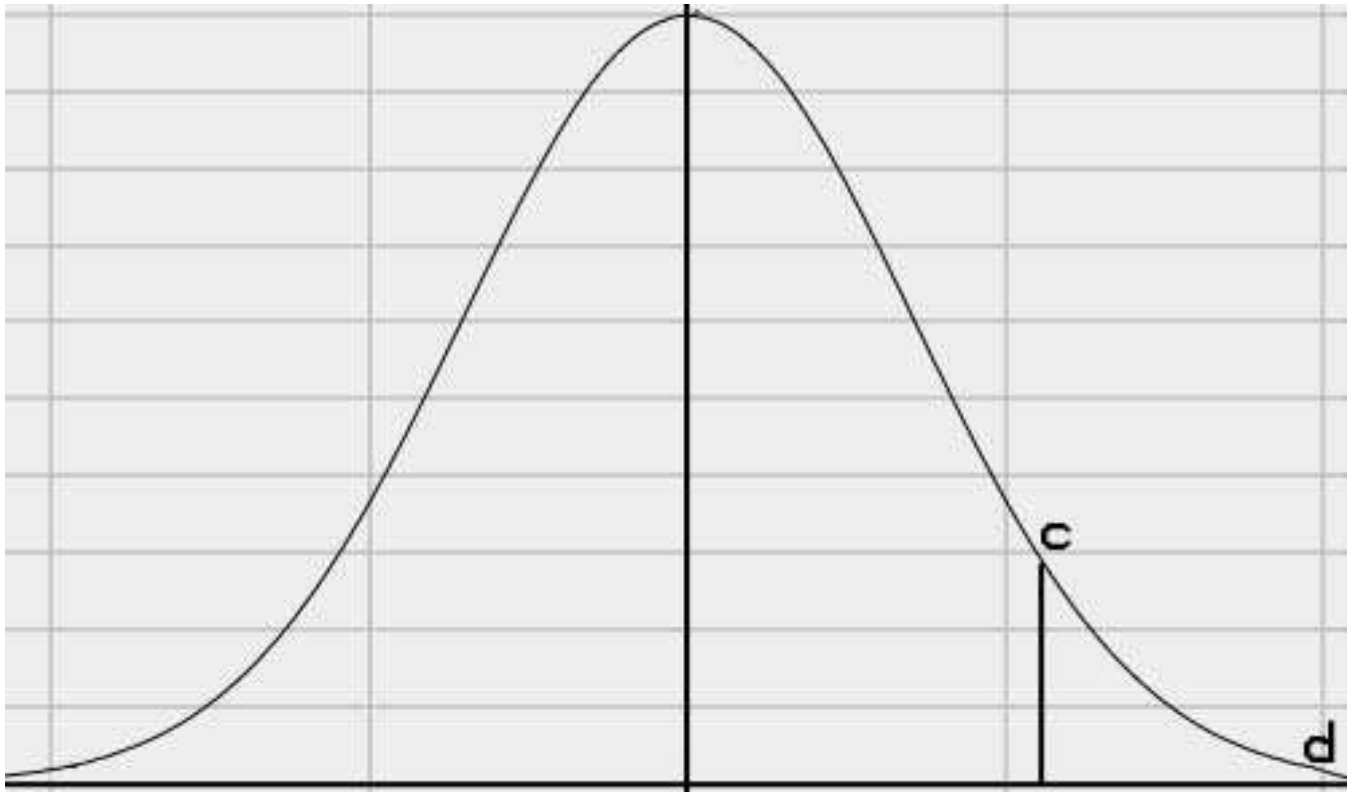}
\caption{The region of interest [c,d] in a loss pdf}
\end{figure}

\noindent In this and the following sections we will also adopt the notation that the first graph in each figure represents a distribution $X_1$, while the second represents a distribution $X_2$. \\

\noindent Although Value at Risk is used by the Basel Committee for determining the capitalization that a bank needs, it turns out that VaR is insufficient for distinguishing between some kinds of different risk characteristics. The following two situations have exactly the same VaR. \\

\begin{figure}[h!]
\centering%
\includegraphics{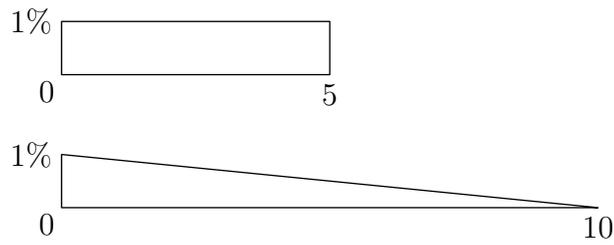}
\caption{Two loss distributions with equal 95\%-VaRs}
\end{figure}

\noindent In each case, the 95\%-Value at Risk is $0$. This is a great misnomer, as there are probabilities of a loss in each case. However, if we consider the notion of Maximum Loss (ML) that is the measure of risk $ML(X)$ that is defined to be the largest loss that a firm could experience with nonzero probability when that firm holds the position $X$ (in the notation of the previous paragraph, we see that $ML(X) = d$), then we can observe that this measure would distinguish between the two situations, as $ML(X_1) = 5$, while $ML(X_2) = 10$. We also note that the 95\%-TCE will identify that the two risk profiles are different, as 95\%-TCE($X_1$) = $\frac{5}{2}$, while 95\%-TCE($X_2$) = $\frac{10}{3}$ .\\

\noindent Since VaR turned out to be insufficient for our risk measuring purposes, maybe one of the two other measures is a better choice. We will start with the Tail Conditional Expectation at the 95\% level. Here we look at two different situations having the same 95\%-Tail Conditional Expectation.\\

\begin{figure}[h!]
\centering%
\includegraphics{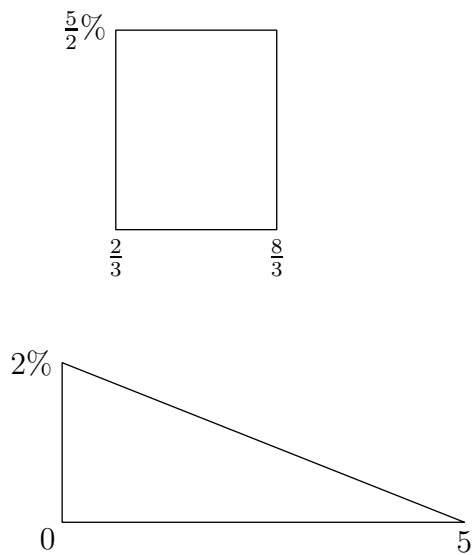}
\caption{Two loss distributions with equal 95\%-TCEs}
\end{figure}

\noindent In Figure 3, both distributions have the same 95\%-Tail Conditional Expectation, as 95\%-TCE($X_1$) = 95\%-TCE($X_2$) = $\frac{5}{3}$, yet there are obviously different risk patterns associated with the two loss distributions. In the first situation, there is uniform probability of loss. In the second case, the probability of loss decreases as losses increase. However, there is the potential for larger loss with the second situation. Each scenario might be preferable to certain institutions for specific situations, but the 95\%-TCE does not alert the institution to any difference in the two different loss patterns. It is notable that both the Maximum Loss and the 95\%-VaR will distinguish between the two situations. 95\%-VaR($X_1$) = $\frac{2}{3}$, while 95\%-VaR($X_2$) = 0. Likewise, $ML(X_1) = \frac{8}{3}$ and $ML(X_2) = 5$.\\

\noindent Because the TCE has not been able to distinguish among different risk situations, we turn to the other measure we have introduced - the Maximum Loss. We will again look at the last 5\% of the probability distribution.\\

\begin{figure}[h!]
\centering%
\includegraphics{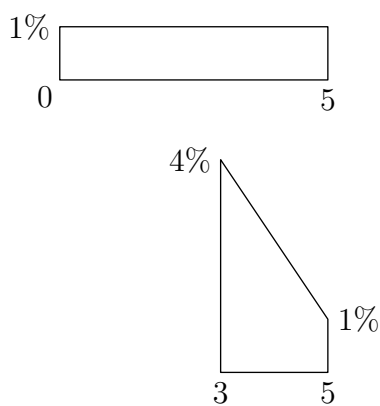}
\caption{Two loss distributions with equal MLs}
\end{figure}

\noindent As we see in Figure 4, the two risk models have the same Maximum Loss as both distributions have no probability occurring after the value of $5$. However, there are vastly different risks involved in choosing one situation over the other. In the first case, there is a simple uniform probability of loss across all the loss levels. In the second case we do not know the probability of losses below 3, as the final 5\% of the distribution does not begin until we get to a loss of 3. However, there is clearly a larger probability of larger losses, tapering off toward the maximum loss. Here both the 95\%-VaR and the 95\%-TCE will determine the difference in the two positions. We first note that 95\%-VaR($X_1$) = $0$ and that the 95\%-VaR($X_2$) = $3$. Also, the 95\%-TCE($X_1$) = $\frac{5}{2}$, while the 95\%-TCE($X_2$) is $3.8$.\\

\noindent From these examples, we should begin to see what each of these risk measures is actually measuring. The 95\% Value at Risk gives the starting point of the worst 5\% of returns under the given model. The 95\% Tail Conditional Expectation tells that point that is the weighted average of all losses in the final 5\% of the probability distribution. And the Maximum Loss shows the worst possible loss for the given risk profile.\\

\noindent In order for us to obtain two distributions with the same 95\%-Value at Risk, we only have to begin our figures at the same value, making sure that we have 5\% of the probability after that common point. For two distributions to have the same 95\%-Tail Conditional Expectation, we must have two geometric figures that have the same weighted average. And for two figures to have the same Maximum Loss, we would require a common point as the maximum value for which both distributions have a nonzero probability.

\subsection*{B. Insufficiency of any two risk measures}

\noindent Since individual measures failed to give a thorough picture of the risk characteristics of a given loss model, maybe the situation will be better if we try using two different risk measures. Having just stated what each risk measure assesses about the risk situation, we will attempt to use measures in pairs to see if they give us a full description of the risk characteristics of a model.\\

\noindent We will start by using both the 95\%-VaR and the 95\%-TCE. Again we are left with a problem. Indeed, the two circumstances shown in Figure 5 have equal 95\%-Values at Risk of $0$ and equal 95\%-Tail Conditional Expectations of $\frac{5}{3}$, yet there are obviously different risk profiles. Again, the first model has uniform loss characteristics. The second has decreasing probability of a larger loss, and it allows for a more severe worst case scenario.\\

\noindent Because we had both the 95\%-VaR and the 95\%-TCE to be equal, we had to meet two conditions. For the VaR condition, we needed both figures to have the same ``starting point'' for the greatest 5\% of losses. Since the first figure is a uniform distribution, it is easy to determine that its weighted center of mass is $\frac{5}{3}$. So our second figure would have to have a weighted center of mass at $\frac{5}{3}$ also, which can be computed by looking at the integral calculation. Note that $ML(X_1) = \frac{10}{3}$, while $ML(X_2) = 5$, so that the measure of Maximum Loss distinguishes between these two distributions.\\

\begin{figure}[h!]
\centering%
\includegraphics{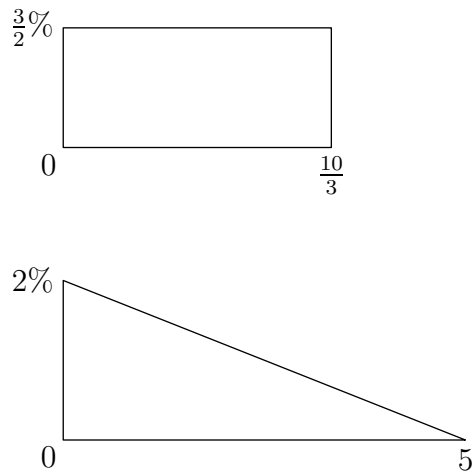}
\caption{Two loss distributions with equal 95\%-VaRs and equal 95\%-TCEs}
\end{figure}

\noindent Now we look at the pair of 95\%-VaR and Maximum Loss, as shown in Figure 6. Once again, we are in the situation where both examples have equality of 95\%-VaR, each of which is $0$. And, for both distributions the Maximum Loss is $5$. Yet, the risk characteristics are once again different. In the second example, there is a larger probability of larger losses, while the first example has uniform probabilities across the entire loss interval. It is notable that the 95\%-TCE here will alert us to the difference in the two situations, as 95\%-TCE($X_1$) = $\frac{5}{2}$, while 95\%-TCE($X_2$) is $\frac{10}{3}$.\\

\begin{figure}[h!]
\centering%
\includegraphics{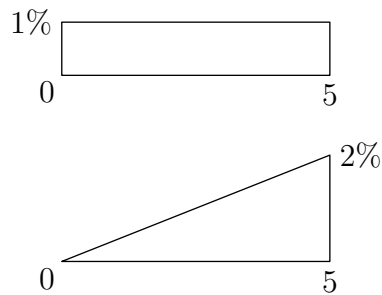}
\caption{Two loss distributions with equal 95\%-VaRs and equal MLs}
\end{figure}

\noindent In order to produce more examples, all we would need is that both have the same starting point of the greatest 5\% of losses and the same ending point of nonzero probabilities. That is, for both distributions we would need a common $c$ and a common $d$ in the notation of the previous section. Above are two such examples that meet these criteria, though many more can be created.\\

\noindent Now we will consider the circumstance of using the 95\%-TCE and the ML as our risk measure pair. In Figure 7, the 95\%-TCEs are equal, as are the MLs. Again, the risk models give different risk characteristics to those who would need to make a choice between the two. We have the uniform loss probability on a longer interval, should we choose the first example. In the second example, the losses increase in probability as the size of losses increases.\\

\begin{figure}[h!]
\centering%
\includegraphics{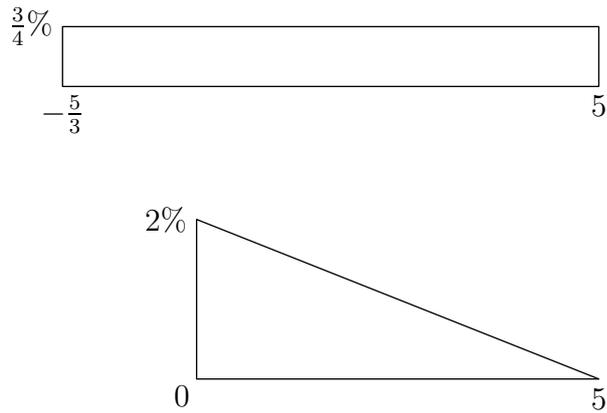}
\caption{Two loss distributions with equal 95\%-TCEs and equal MLs}
\end{figure}

\noindent Here, we again have two criteria that have to be met. In order for the 95\%-TCEs to be equal, we need to have the same weighted centers of mass. Again, these can be determined by doing the integral calculations that were previously described. For the MLs to be equal, we again need both distributions to have the same ``ending point.''\\

\noindent For both of the distributions we have that the 95\%-TCEs are $\frac{5}{3}$, while the MLs are both $5$. Thus, our examples show that any two of our risk measures are insufficient to distinguish among loss distributions. We notice that 95\%-VaR($X_1$) = $-\frac{5}{3}$, while 95\%-VaR($X_2$) = 0, so that the third measure differentiates between the two risk profiles. \\

\subsection*{C. Insufficiency of all three measures}

\noindent With each pair of risk measures proving insufficient to determine the entire risk loss pattern, we look to the potential that three measures might work. Our four examples in Figure 8 have equal 95\%-VaRs, equal 95\%-TCEs, and equal MLs, yet the risk patterns are clearly different. The first is once again a uniform loss pattern. The second example shows increasing probabilities of losses and then decreasing, while the third example does just the opposite. The fourth example shows that this pattern of increasing and then decreasing probabilities (or likewise decreasing and then increasing probabilities) could be appropriately extended to get an infinite family of examples which have equality of all three of our measures. \\

\begin{figure}[h!]
\centering%
\includegraphics{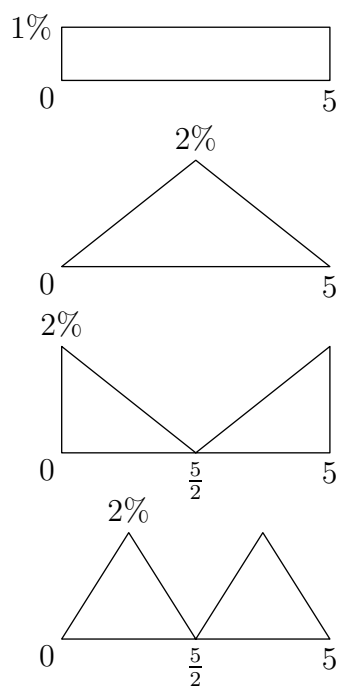}
\caption{Four loss distributions with equal 95\%-VaRs, 95\%-TCEs and MLs}
\end{figure}

\noindent In order for our examples to have equivalence of all three measures, we have to meet three criteria. We need the same ``starting points'' of the worst 5\% of retruns in order to make the VaRs equal. We require identical ``ending points'' for the MLs to be equal. Since the uniform distribution has symmetry, it will have a 95\%-TCE of $\frac{5}{2}$. Thus, any distribution with an equivalent 95\%-TCE (and the same starting and ending points) must also be symmetric in the greatest 5\% of its losses.\\

\noindent So our three measures do not distinguish among the potential scenarios that we might face. Obviously, something more is needed.

\subsection*{D. Extending to different probability levels}

\noindent We have chosen to use the 95\% level for our risk measures, but there is nothing special about the 95\% level. Our examples would be just as valid if we substituted 90\% or 99\% or any other number in all of our measures and then scaled identically in all of the corresponding diagrams. However, this concept of considering different risk levels might also lead us to enough risk measures so that we can determine the risk profile from information about the different risk measures.\\

\noindent So now we will try using a vector of measures consisting of the 95\%-VaR, the 99\%-VaR, the 95\%-TCE, 99\%-TCE, and the ML. The different risk profiles demonstrated in Figure 8 are quite odd-looking, but they all have identical 95\%-VaRs, 99\%-VaRs, 95\%-TCEs, 99\%-TCEs, and MLs. \\

\begin{figure}[h!]
\centering%
\includegraphics{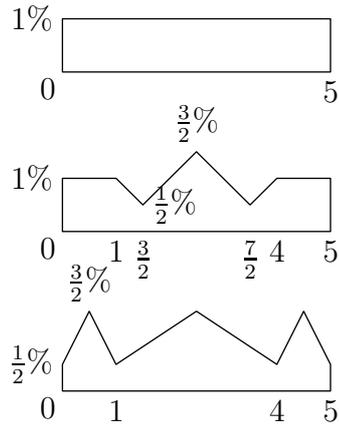}
\caption{Using five risk measures}
\end{figure}

\noindent The three distributions all have the necessary geometric properties for equality of all five measures. If we wanted to form a larger family of examples with the same values for all five measures, we would need our new distributions to have the same ``ending points'' in order to match up the MLs. We would need them to have the same ``starting points'' for both the greatest 5\% of losses and the greatest 1\% of losses in order to have equal 95\%-VaRs and equal 99\%-VaRs, respectively. And we would need symmetry in the greatest 5\% of losses and in the greatest 1\% of losses in order to have 95\%-TCEs and 99\%-TCEs, respectively, equal to the 95\%-TCE and the 99\%-TCE of the uniform distribution. Any figure meeting all these criteria would also be indistinguishable from the first five when viewed purely by looking at the values in our vector of five measures. Clearly several different profiles can share all five measures.

\section*{V. Conclusion and Future Directions}

\noindent We suggest that banks be required to submit multiple risk measures to regulatory bodies. We have shown that individual risk measures and combinations of measures can lead to ambiguity in trying to determine the actual loss distribution. Even the reporting of five risk measures can theoretically lead to an infinite family of possibilities for the actual loss distribution.\\

\noindent This is an interesting theoretical result that must be pursued on actual loss data. A paper that looks at the ability to differentiate between different types of common loss distributions is being written with my student, Cole Arendt. I am presently considering the question of how many risk measures are needed if we vary the timeframes on the historical data along with the level and the specific measures. Finally, another paper that looks at the predictive ability of risk measures is also in the works.\\

\end{document}